\documentclass[preprint,showkeys,aps,prb]{revtex4}
\usepackage[dvips]{graphicx}
\begin{document}

\title{An Appreciation of Berni Julian Alder}

\author{
William Graham Hoover                      \\
Ruby Valley Research Institute             \\
Highway Contract 60, Box 601               \\
Ruby Valley, Nevada 89833                  \\
}

\date{\today}


\vspace{0.1cm}

\begin{abstract}
Berni Alder profoundly influenced my research career at the Livermore Laboratory and the Davis
Campus' Teller Tech, beginning in 1962 and lasting for over fifty years.  I very much appreciate
the opportunity provided by his Ninetieth Birthday Celebration to review some of the many high
spots along the way.

\end{abstract}

\maketitle

\section{Ann Arbor Revisited and Durham explored}
My Father, Edgar Malone Hoover, Junior, taught economics at Harvard ( Ph D 1928 ) and the University
of Michigan until World War II brought him to Washington for work with the National Resources Planning
Board, the Office of Price Administration, and the Office of Strategic Services. After half a dozen
of those years in Washington, followed by a Chemistry major at Oberlin College ( AB 1958 ) I returned
to Ann Arbor for a Ph D in Chemical Physics, 1958-1961 .  Three of the scientific highlights of those
years were [1] a short course in FORTRAN ( a three-hour lecture, taught in a single evening ); [2]
George Uhlenbeck's lectures on {\it Gastheorie}, delivered while holding his musty notes at arm's
length; and [3] Andrew De Rocco's course on statistical mechanics.  I was specially inspired by the
computer-generated pictures in Berni and Tom Wainwright's ``Molecular Motions'', published in the 1959
Scientific American\cite{b1}. See {\bf Figure 1}. When I saw their work I wanted to make some of these
manybody dynamics pictures myself.

\begin{figure}
\includegraphics[width=4.5in,angle=+90]{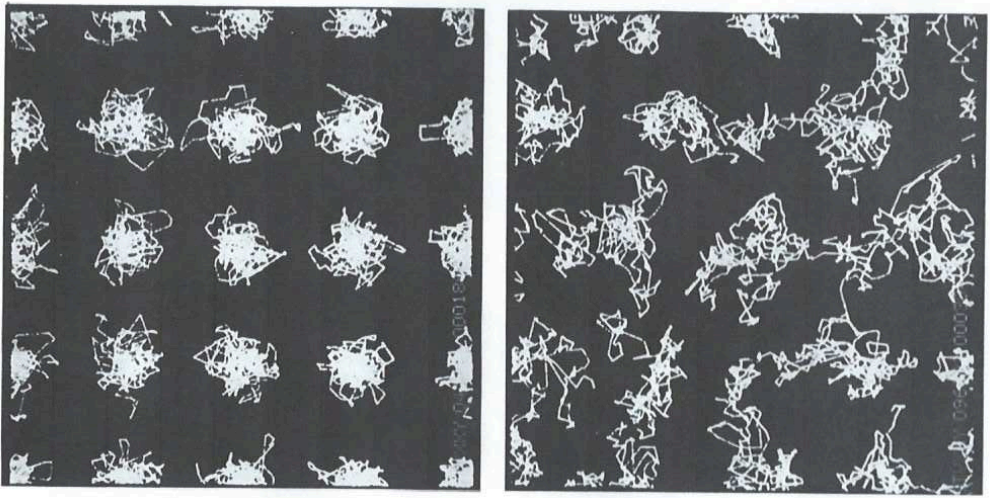}
\caption{
Particle trajectories of 32 hard spheres with spatially periodic boundary conditions. At densities
less than two-thirds of close packing only the fluid phase is stable.  From Reference 1.\\
}
\end{figure}

In the early 1960s the main theoretical route to equations of state was through integral equations for the
pair distribution function.  This elegant and absorbing approach was soon made thoroughly obsolete
by the two simulation techniques of molecular dynamics and Monte Carlo.  In 1961-1962 I spent a year
in  postdoctoral work at Duke with one of John Kirkwood's students, Jacques Poirier. The result was
an improved understanding of integral equations and the Mayers' virial series.  At Duke I shared an
office with Jacques' student, John Nelson Shaw, whose Ph D project was the development of a
one-component-plasma Monte Carlo code. Each of the variables in John's computer program was named for
a member of his Family.  The code was used soon after by Steve Brush, Harry Sahlin, and Edward Teller
at the Livermore Laboratory. Computing was slow in those days. Moore's 1965 Law was not yet known.
Automatic equation-of-state calculations, though soon to be commonplace, were still a few years away
in the unforeseeable future.

\section{Los Alamos and Livermore}
When it came time to find a ``real job'' I was still motivated by the Alder-Wainwright Scientific
American article and applied to both Livermore and Los Alamos, where the best computers were.  I had
interview talks with Bill Wood at Los Alamos and Berni at Livermore.  Both places were appealing
with rather different physical environments but wonderful opportunities for computational research.
The higher salary offer ( though with much shorter vacations ) brought me to Livermore, providing
the chance to do real simulations rather than follow the not-so-reliable and not-so-simple
integral-equation and virial-series paths.

My first California publication, with Berni and Tom\cite{b2}, showed that two hard disks, with periodic
boundary conditions, give a van der Waals' pressure-volume loop within a few percent of the large-system
transition that had been Berni's interest ever since his doctoral work with Kirkwood.  See {\bf Figure 2}.
Research at Livermore in the early 1960s was a joy, free from the need to apply for grants or to write
progress reports.  My efforts were strengthened by stimulating  collaborations with Francis
Ree, Tom Wainwright, and Berni. Often our relatively-long research days were divided up
by dinner discussions at Livermore's Yin Yin restaurant ( 1960 - present ).  In those days abalone
was still on the menu.

\begin{figure}
\includegraphics[width=4.5in,angle = +90.]{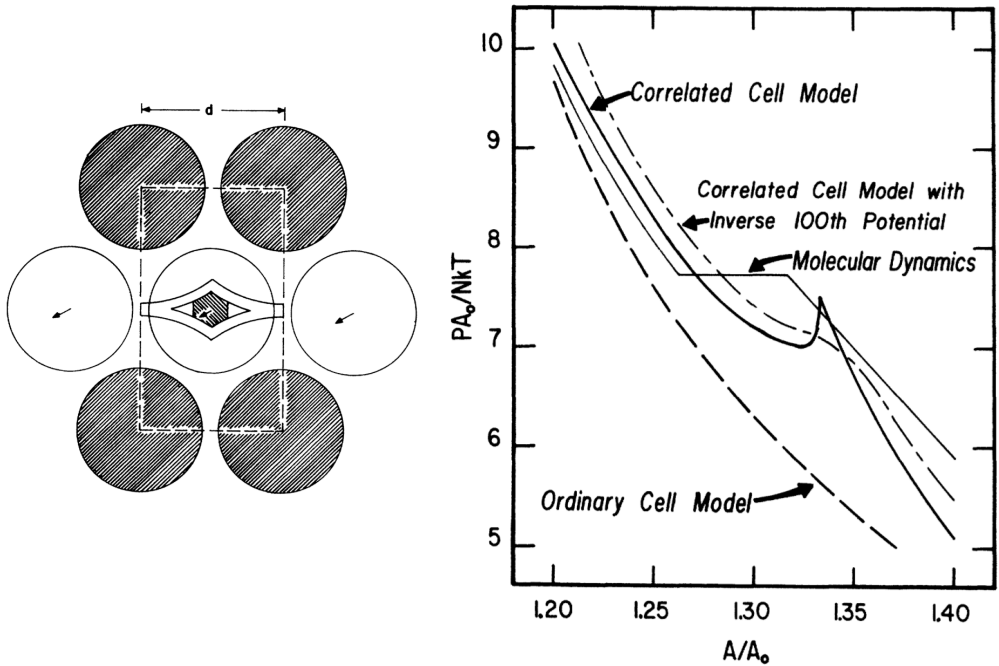}
\caption{
In the left view a row of hard disks moves cooperatively, expanding the ``free volume''
explored by about one third and allowing a phase transition as that periodic row slips past its
neighbors.  Without this cooperative motion the ``ordinary'' cell model predicts a pressure somewhat
lower than the results of manybody simulations shown at the right. From Reference 2.\\
}
\end{figure}

\section{Work with Francis Ree}
Francis Ree was a student of Henry Eyring's at the University of Utah, mathematically gifted and a
perfect coworker.  He and I had the idea to implement Kirkwood's ``communal-entropy'' model into
Monte Carlo simulations measuring hard-disk and hard-sphere entropies. Kirkwood's idea was that fluid
phases enjoyed an additional shared entropy $Nk$ as a consequence of indistinguishability.  This
shared entropy was identified with an extra $e^N$ in fluid partition functions.  With Edward Teller's
permission to use the considerable computer time entailed Francis and I measured the communal entropy for both
disks and spheres. We found that the ``extra'' fluid entropy varies slowly with density rather than
appearing suddenly at melting. We used quantitative measurements to compute precisely the density
dependence of the entropy difference between the fluid and solid phases.  Our communal-entropy
work\cite{b3}, along with  solid-phase investigations carried out with Berni and David Young\cite{b4},
led to accurate locations of the melting transitions for both disks and spheres.

By the late 1960s Francis and I had had enough virial series and entropy work on the hard-disk and hard-sphere
problems. We sought out new directions.  Francis enjoyed phase diagram work while I, now working for
Russ Duff, pursued nonequilibrium studies. In 1967 I was the very last one of seven authors on a
shockwave paper presented by Russ in Paris\cite{b5}.  This work documented progress toward simulating
shockwaves with {\it continuous-potential} molecular dynamics. The feasibility of such simulations
had been established by Enrico Fermi ( at Los Alamos ), George Vineyard ( at Brookhaven ), as well as
Aneesur Rahman ( at Argonne ). I had missed the Fermi and Vineyard work.  But Rahman's later work did
get my attention\cite{b6} and I implemented his predictor-corrector algorithm.  

Brad Holian, Bill Moran, Galen Straub, and I\cite{b7} revisited shockwave simulations in the 1980s,
using the much more efficient leapfrog algorithm.  That work focused on the strong compression of liquid
states, as is illustrated in {\bf Figure 3}. Decades later, by then converted to Runge-Kutta, and
working with Paco Uribe and my Wife Carol\cite{b8}, we studied more detailed models of the shock process.
These models included tensor temperature, with $T_{xx} >> T_{yy} = T_{zz}$ .  We also found and included
time delays between the fluxes of momentum and energy and the velocity and temperature gradients
driving these fluxes.  {\bf Figure 4} shows the pressure tensor and the heat-flux vector for a simple model
including these effects.  Such a model can describe numerical molecular dynamics data rather
well\cite{b8}.\\

\begin{figure}
\includegraphics[width=4.5in,angle = +90.]{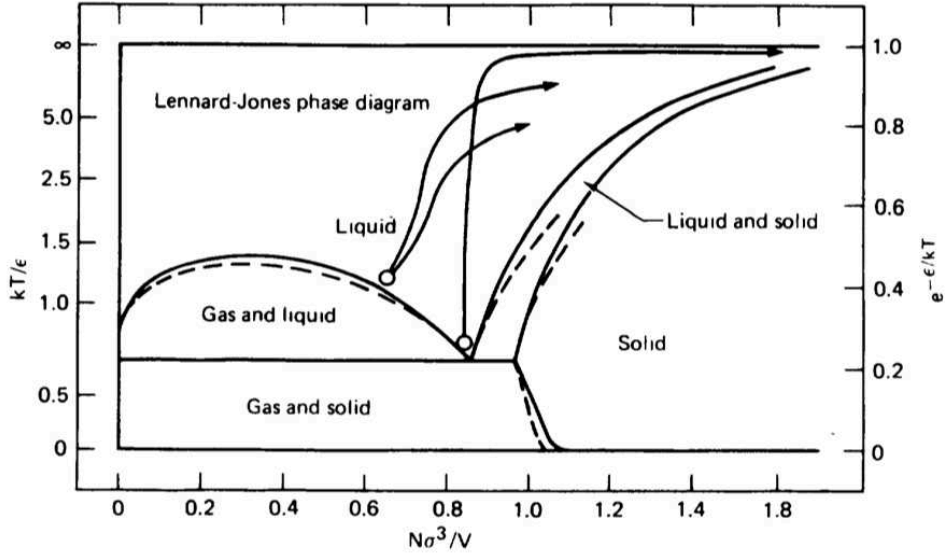}
\caption{
The temperature-density trajectories of three shockwave simulations starting in the liquid phase
and ending up as highly-compressed hot gas.  The dashed lines represent the phase diagram of liquid
argon. From Reference 7.
}
\end{figure}

\begin{figure}
\includegraphics[width=4.5in,angle = +90.]{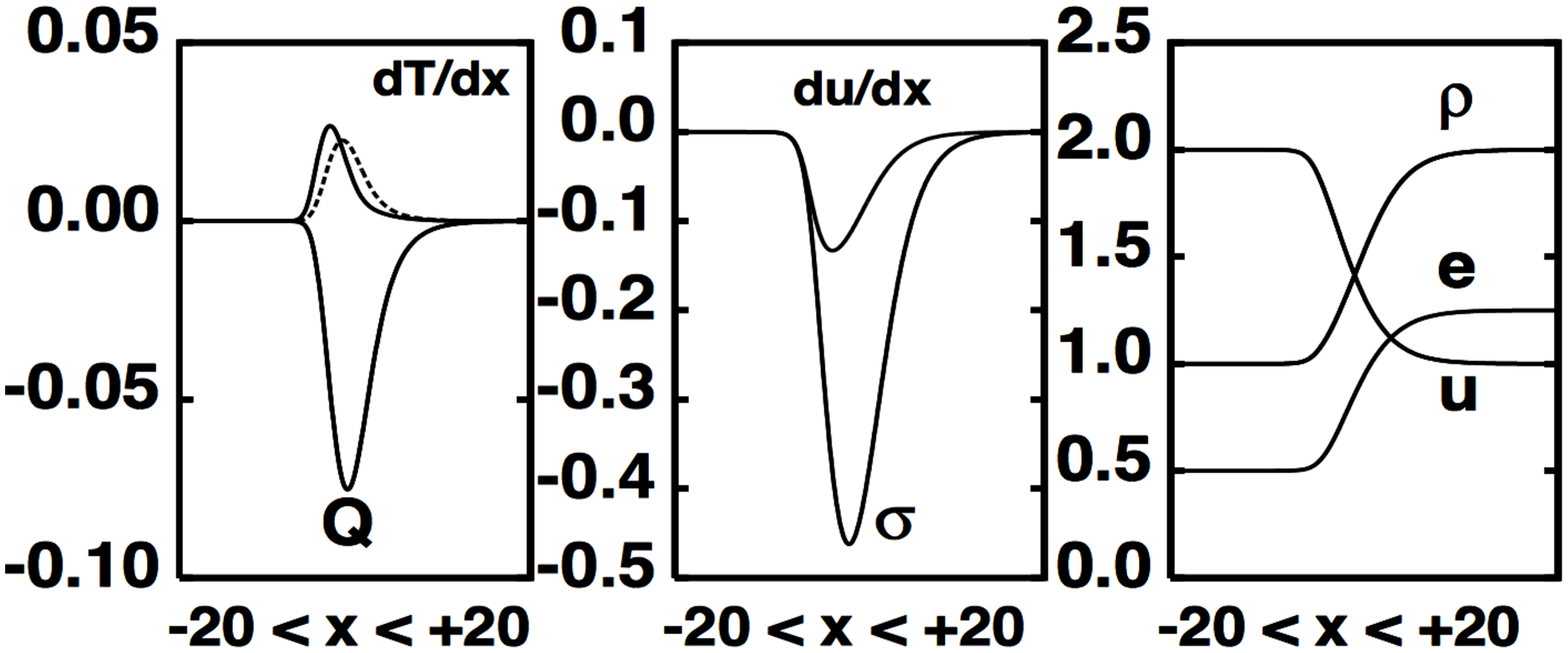}
\caption{
The time delay between stress $\sigma$ and strain rate $(du/dx)$ and between heat flux $Q$ and
the temperature gradients $(dT/dx)$ ( longitudinal and transverse ) are shown for continuum
simulations of moderately-dense shockwaves. These models describe the results of molecular dynamics
simulations quite well. The figure is taken from Reference 8. \\
}
\end{figure}

\section{Bill Ashurst and Nonequilibrium Molecular Dynamics}
Once several workers (Hans Andersen, John Barker, Frank Canfield, David Chandler, Doug Henderson, Ali Mansoori, Jay Rasaiah,
George Stell, and John Weeks) had developed a hard-sphere-based  perturbation theory of equilibrium
liquid states {\it nonequilibrium} simulations opened up as a more promising source of new ideas. 
In 1971 Berni had helped me into a part-time Professorship at U C Davis through Edward Teller's
Department of Applied Science.  My first Ph D student there, Bill Ashurst, from Sandia's Livermore
Laboratory, became interested in modeling shear and heat flows directly, with {\it thermostated}
``nonequilibrium molecular dynamics''\cite{b9}.

Bill used differential feedback to maintain constant boundary velocities and temperatures.  Some of
his transport coefficients disagreed with those computed from Green-Kubo linear-response theory ( but
with factor-of-two errors ) by our French colleagues, Levesque, Verlet, and  K\"urkijarvi\cite{b10}.
Thanks to a grant from the Academy of Applied Science ( Concord, New Hampshire ), made possible
through my U C Davis position, I was able to finance summer research projects with bright high
school students. The same grant enabled considerable foreign travel so that I could exchange ideas
with the many researchers gathered together in France. Our interests in common with these researchers
led to numerous trips to Paris and Orsay to participate in Carl Moser's simulation workshops.
The European contacts were stimulating and pleasant and led to a productive sabbatical in Wien
followed up by a thirty-year collaboration with Harald Posch and his colleagues.

Some of the puzzles that arose from this 70s-80s-era work remain today.  For example, the effective
viscosity measured in a strong shockwave exceeds the small-strainrate Newtonian one by tens of
percent.  At the same time a steady {\it homogeneous} shear flow at the same strain rate exhibits
a {\it reduced} viscosity, also by tens of percent. These opposite nonlinear effects indicate that
there is still much to learn about nonlinear transport.

\section{Sabbatical Research in Australia, Austria, and Japan}
My academic connection to the University of California at Davis made it possible to go on
sabbaticals --- Canberra ( 1977-1978 ), Wien ( 1985 ), and Yokohama ( 1989-1990 ) .  The year in
Australia enabled my son Nathan, just graduated from Livermore High, to work with me, exploring another
of Berni's interests, hard-disk and sphere ``free volumes''. Kenton Hanson did the three-dimensional
work in Berkeley.  Nathan and I discovered a percolation transition for hard disks, where the free volume changes
from extensive to intensive, at one-fourth of the hard-disk close-packing density\cite{b11}.  See
{\bf Figure 5} . It was interesting to see that solid-phase free volumes are larger than the fluid
ones at the same density.

In Wien I worked with Karl Kratky, Harald Posch, and Franz Vesely while lecturing and writing
my first book, ``Molecular Dynamics''.  The year in Japan, arranged by Shuichi Nos\'e, was shared
with my new wife Carol, resulting in a longer sequel book, ``Computational Statistical
Mechanics'' with inspiration and cover art from Shuichi ( Yokohama ) and Harald Posch
( Wien ).  See {\bf Figure 6} .  Nos\'e's work changed my outlook on computer simulation.  Let us
look at some of the details.\\

\begin{figure}
\includegraphics[width=4.5in]{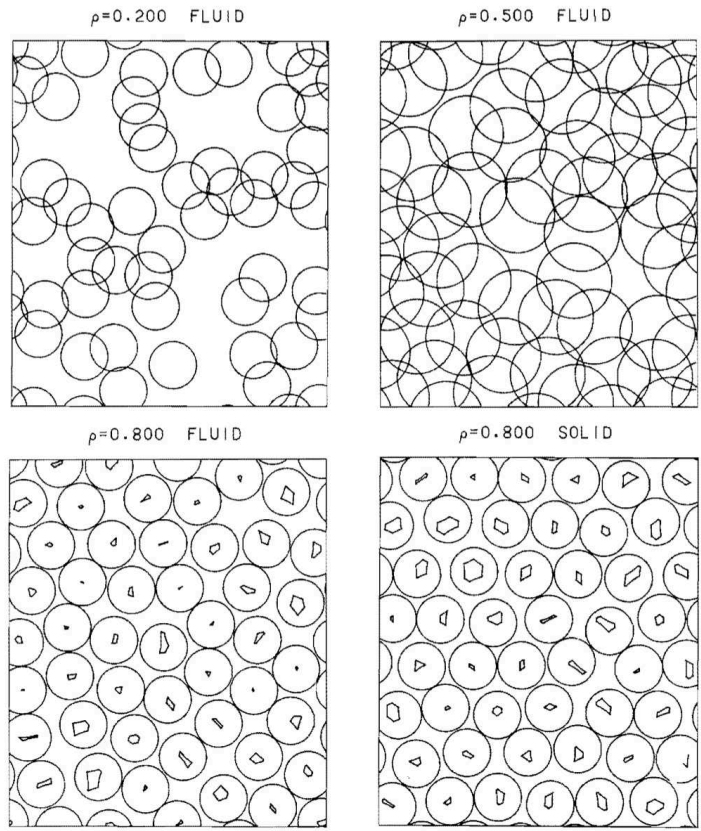}
\caption{
The top row shows exclusion disks which the centers of other disks cannot penetrate.
The lower row shows the particles themselves and their free volumes at a density near that of
the fluid-solid transition.  The pictures in the top row make it plausible that there is a
percolation transition (from extensive free volumes to intensive) between the densities
shown.  In Reference 11 numerical work shows that the transition density is close to one-fourth
the close-packed density.\\
}
\end{figure}

\begin{figure}
\includegraphics[width=4.5in,angle=+90]{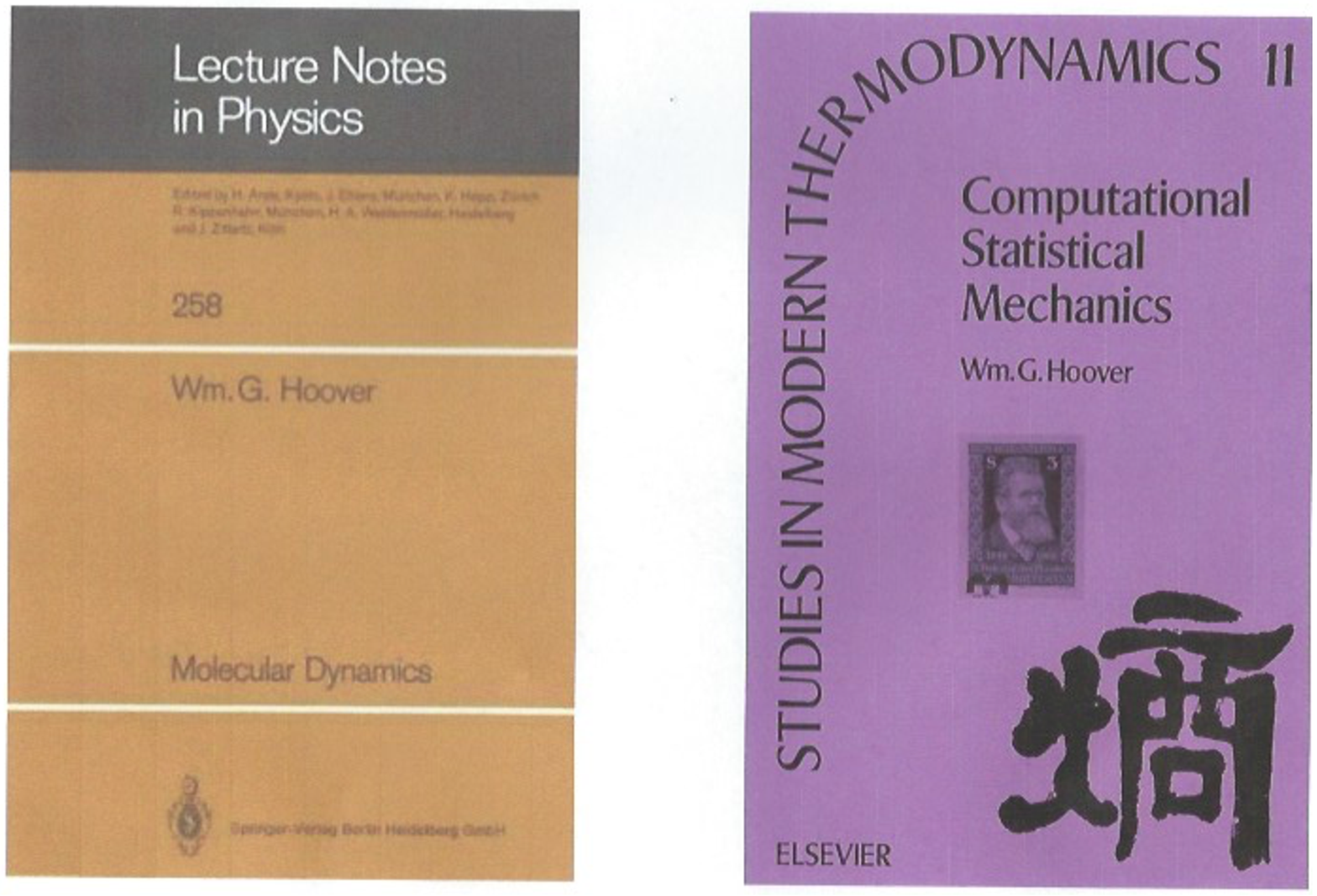}
\caption{
Two books written during sabbaticals in Austria and Japan.  The two-part Japanese character
combining ``heat'' and ``divide'' represents ``entropy''.  Note Boltzmann's presence here.
}
\end{figure}

\noindent
\section{Shuichi Nos\'e and Keio University}
During Orwell's ``1984'' I came across two amazing papers\cite{b12,b13} in the Library of Lawrence's
Livermore Laboratory.  They were written by a then-unknown Japanese postdoc, Shuichi Nos\'e, who was working
in Canada with Mike Klein.  The title of one of Shuichi's papers\cite{b12} describes the gist of his
work, ``A Molecular Dynamics Method for Simulations in the Canonical Ensemble''.  In 1984 this title's
concept seemed to me completely paradoxical.  For me ``Molecular Dynamics'' had generally meant the
{\it micro}canonical constant-energy ensemble, not the very different constant-temperature canonical one.
Although Bill Ashurst and I had long carried out ``isokinetic'' simulations, for over a decade, the notion
of a ``canonical'' dynamics made no sense to me.  I set out to meet Nos\'e at an upcoming workshop in
France and was lucky enough to find him, completely by accident, in Paris' Orly train station days prior
to the workshop's start.

Nos\'e's novel thermostat ideas took me a couple of weeks to digest back then in 1984, even after
several hours of conversation before and during the  CECAM [ European Center for Atomic and
Molecular Calculations ] workshop\cite{b14}.  Following up Nos\'e's new ideas, applied to the
harmonic oscillator, revealed a troubling aspect of his new dynamics.  The results for long-time
averages depended strongly
on the initial conditions.  A harmonic oscillator, in one dimension and with unit force constant
and mass, was the simplest illustration. I could see that his four motion equations, which included
a completely novel ``time-scaling'' variable $s$, its conjugate momentum $\zeta$, and a ``number of
degrees of freedom'' $\# = 2$ , one for $(qp)$ and one for $(s\zeta)$ :
$$
\{ \ \dot q = (p/s^2) \ ; \ \dot p = -q \ ; \ \dot s = \zeta \ ;
\ \dot \zeta = (p^2/s^3) - (\#/s) \ \} \ [ \ {\rm N} \ ] \ ,
$$
could be replaced by an equivalent but simpler and less stiff set of just {\it three} equations, with $s$
absent and $\# = 1$ :
$$
\{ \ \dot q = p \ ; \ \dot p = -q - \zeta p \ ; \ \dot \zeta = p^2 - \# \ \} \ [ \ {\rm NH} \ ] \ .
$$
Solutions of these three ``Nos\'e-Hoover'' equations give $\zeta(q)$ trajectories {\it identical}
to those from the four ``Nos\'e'' equations provided that (1) the $\#$s match, (2) the initial
value of $s$ is unity, and (3) the other initial values match: $(qp\zeta)_{\rm N} =
(qp\zeta)_{\rm NH}$ .
And it was true, as Nos\'e had pointed out in his papers, that the three equations were likewise
consistent with Gibbs' canonical distribution, along with an additional Gaussian distribution for
the momentum variable $\zeta$ :
$$
f(q,p,\zeta) \propto e^{-q^2/2}e^{-p^2/2}e^{-\zeta^2/2} \ .
$$
In fact, by starting with Gibbs' distribution and working backward using Liouville's phase-space
flow equation, I could obtain Nos\'e's ``time-scaled'' dynamics without any consideration of
time scaling at all !

Besides the mysterious time-scaling there was still that troubling fly in the ointment. The thermostated
harmonic-oscillator dynamics doesn't actually generate {\it all} of Gibbs' distribution.  Instead six
percent of its Gaussian measure occupies a ``chaotic sea'' in which nearby trajectories separate
from one another with a positive exponential growth rate proportional to $e^{+\lambda t}$ where
$\lambda = 0.0139$ is the system's largest ``Lyapunov exponent''\cite{b15}.  All of the remaining 94 percent
of the Gaussian distribution is occupied by an infinite set of periodic orbits and their tori.  Each such
periodic orbit is surrounded by concentric
stable toroidal orbits with {\it vanishing} Lyapunov exponents, $\lambda = 0$ .  The complexity of
this distribution can be visualized in the $(0p\zeta)$, $(q 0 \zeta)$, and $(qp0)$ cross sections of
{\bf Figure 7}.  Understanding all of this new information evolved gradually, over a period of years
rather than days.\\

\begin{figure}
\includegraphics[width=4.5in,angle=-90.]{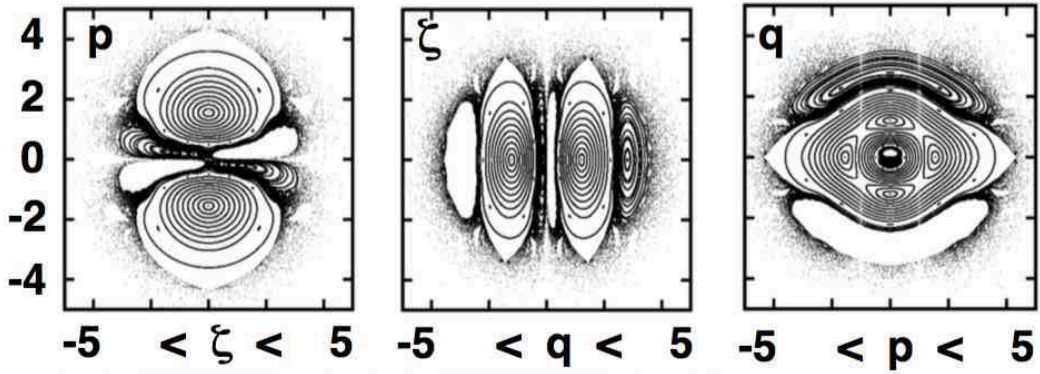}
\caption{
Cross sections of the chaotic sea with a selection of toroidal closed-curve orbits for the
equilibrium Nos\'e-Hoover oscillator with $T = 1$ .  From Reference 15.\\
}
\end{figure}

Ultimately, meeting Shuichi Nos\'e in Paris in 1984 led me into many new research directions running
the gamut from one-body chaos to many-body hydrodynamics, combining ideas from dynamical systems
and chaos theory with continuum mechanics and molecular dynamics.  In turn this led to a strong
and prolific collaboration ( 50 joint papers ) with Harald Posch at Boltzmann's University on
Boltzmanngasse in Wien.  By this time I had moved away from Berni's research interests, which had
become mainly quantum mechanical.  My own work with Harald and later with Carol involved thermostats
and the connections linking continuum mechanics to molecular dynamics and dynamical systems theory.

Many of these projects involved Brad Holian, a gifted student of Berni's and a real enthusiast for
{\it Many}-body molecular dynamics.  At the other extreme, a single degree of freedom, Brad and I were able
to show that the troubling one-dimensional harmonic oscillator, with not only its second velocity
moment controlled, but also its fourth, {\it is} ``ergodic''\cite{b16}.  That is, with both velocity moments
controlled time reversibly, this four-equation model provides {\it all} of Gibbs' canonical distribution
for the oscillator as well as Gaussian distributions for the two thermostat variables $\zeta$ and $\xi$ :
$$
\{ \ \dot q = p \ ; \ \dot p = -q - \zeta p - \xi (p^3/T) \ ; \ \dot \zeta = (p^2/T) - 1 
\ ; \ \dot \xi = (p^4/T^2) - 3(p^2/T) \ \} 
$$
$$
\longrightarrow (2\pi)^2Tf(q,p,\zeta,\xi) \equiv e^{-q^2/2T}e^{-p^2/2T}e^{-\zeta^2/2}e^{-\xi^2/2} \ .
$$

The idea of formulating a canonical-ensemble dynamics ( giving a Gaussian velocity distribution
defined by its kinetic temperature $\langle \ p^2 \ \rangle$ 
rather than constant energy ) was an outgrowth of Shuichi Nos\'e's 1984 Magicianship\cite{b12,b13}.
Carol and I had married in 1988 in preparation for a very productive year (1989-1990 ) together at
Nos\'e's Keio University. The working conditions in Japan were extremely pleasant, walking to work
at Keio University's Hiyoshi campus, and with no real duties other than speaking at a few conferences
during the year. We collaborated with Tony De Groot who had built a CRAY-speed computer back at
Livermore with a transputer budget of only \$30,000.  With many colleagues' help we were able to
simulate plastic flow with millions of degrees of freedom in reasonable clock times with Tony's
machine.  See {\bf Figure 8}\cite{b17}.\\

\begin{figure}
\includegraphics[width=4.5in,angle=+90.]{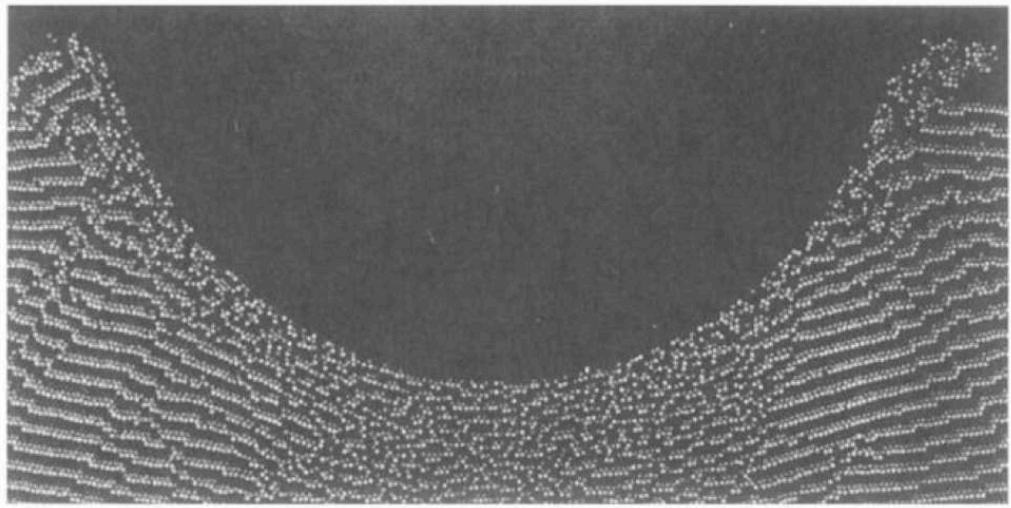}
\caption{
Plastic flow around an indentor with more than a million atoms. A simulation carried out on
Tony De Groot's low-cost transputer computer, the ``SPRINT''. From Reference 17.\\
}
\end{figure}

\section{Recent Studies of Ergodicity and Lyapunov Instability}

During 2014-2015 Carol and I have been enjoying a very fruitful collaboration with Clint Sprott ( Wisconsin )
and Puneet Patra ( a Ph D student of Baidurya Bhattacharya's in Kharagpur )\cite{b18,b19,b20}. We
studied the ergodicity and Lyapunov instability of oscillators exposed to a temperature gradient,
$T(q) = 1 + \epsilon \tanh(q)$. Over the course of a year we came upon {\it several} ergodic
oscillator models with only {\it three} motion equations, rather than four. This was a pleasant
surprise. The analysis of three-dimensional topology rather than four- is a tremendous simplification.

The simplest of all these three-dimensional models incorporates ``weak'' (integrated) control of both
$p^2$ and $p^4$ :
$$
\dot q = p \ ; \ \dot p = -q - 0.05\zeta p - 0.32\zeta (p^3/T) \ ; \
\dot \zeta = 0.05[ \ (p^2/T) - 1 \ ] + 0.32[ \ (p^4/T^2) - 3(p^2/T) \ ] \ .
$$
For {\it any} initial condition the Gaussian velocity distribution results.  And now the chaotic sea
embraces the entire three-dimensional phase space.

{\bf Figure 9} displays the sign ( red positive and green negative ) of the local Lyapunov exponent
just at the moment that the equilibrium $(qp\zeta)$ trajectory crosses the phase plane $\zeta = 0$ . The
``local'' Lyapunov exponent describes the instantaneous growth or decay rate of a small displacement in the
neighborhood of a ``reference trajectory''. In the isothermal constant-$T$ case an amazing consequence
of the three thermostated equations of motion is a simple three-dimensional Gaussian distribution
generated by an incredibly-complicated Lyapunov-unstable, but time-reversible, one-dimensional
trajectory.  {\bf Figure 9} conceals a surprise.  If a $(qp\zeta)$ oscillator is viewed in a mirror
perpendicular to the $q$ axis both $q$ and $p$ change sign, corresponding to inversion of the cross
section through the origin at its center. Diverging trajectories viewed in a mirror likewise diverge.
There is also a {\it missing} symmetry.  One would think that running a trajectory backwards, with $q$
unchanged and $p$ reversed in sign would replace divergence by convergence so that red in the lower
half plane would correspond to green above, and {\it vice versa} .  This symmetry, though fully
consistent with the time-reversible motion equations, is absent.  The reason is that the tendency
toward divergence or convergence of two trajectories, constrained by a tether, {\it can only depend
upon the past}, and {\it not the future}.
This relatively subtle distinction can be clarified by considering the {\it nonequilibrium} case where
$T$ {\it varies} with the coordinate $q$ .  We turn to that case next. 

In the {\it nonequilibrium} case with temperature $T(q) = 1 + \epsilon \tanh(q)$ -- so that the maximum
temperature gradient is $\epsilon$ -- the equations remain time-reversible.  Reversing the direction of
time ( as well as the signs of $p$ and $\zeta$ ) could be expected to reverse the sign of the Lyapunov
exponent.  {\bf Figures 9 and 10} show that this natural expectation is unwarranted and wrong.  The top and
bottom halves of these figures are {\it not} simple mirror images.  Evidently the time-reversiblity of all
the equations is misleading.  Let us consider this observed symmetry breaking in more detail.\\

\begin{figure}
\includegraphics[width=4.5in,angle=+90.]{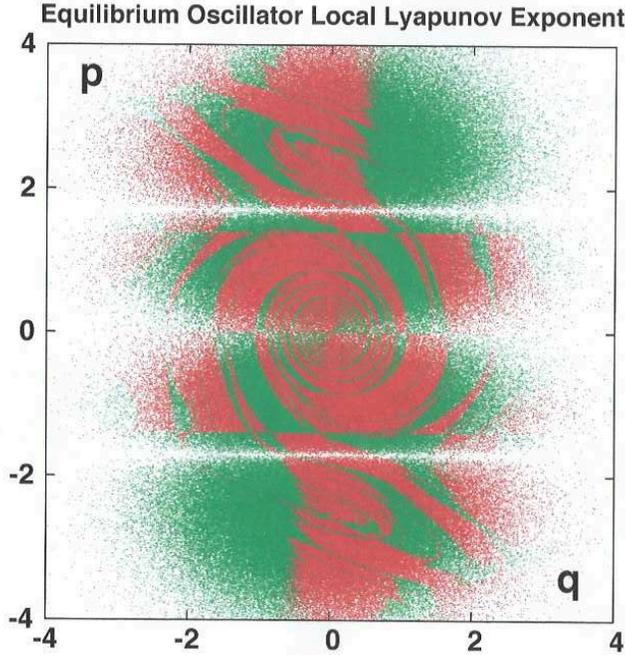}
\caption{
Local Lyapunov exponents (red for positive and green for negative) for the three-equation
equilibrium ergodic oscillator described in the text.  The imposed constant-temperature profile is
$T(q) = 1$ . Notice that the color for $(\pm q,\pm p)$ is identical to that for $(\mp q,\mp p)$
corresponding to the mirror ( inversion ) symmetry of the problem.  This symmetry also corresponds
to 180 degree rotation of the $(qp0)$ cross section.  Time-reversal symmetry of the
Lyapunov exponent is {\it absent}.
}
\end{figure}

\section{Symmetry Breaking in Time-Reversible Flows}

From the conceptual standpoint an interesting paradoxical aspect of ( irreversible )
nonequilibrium flows is the time reversibility of their underlying motion equations.  This
constrast with ``real life'' motivates Loschmidt's and Zerm\'elo's Paradoxes.  Loschmidt's
is that forward and backward movies of flows {\it satisfying} the irreversible Second Law
of Thermodynamics are described by exactly the same time-reversible motion equations
in both time directions.  Zerm\'elo's is that any initial phase-space state, no matter
how unlikely or odd, will eventually recur due to the bounded nature of the constant-energy
phase space.  Reversal and recurrence both seem to violate the Second Law.

By looking at small nonequilibrium systems [ such as a harmonic oscillator exposed to a
temperature gradient, $T(q) = 1 + \epsilon \tanh(q)$ ], we found that the phase-space
description of such systems is invariably a unidirectional flow ``from'' a fractal
``repellor'' ``to'' a ``strange attractor''.  The attractor and repellor correspond to
velocity mirror images.  The motion forward in time is attractive, with diminishing phase
volume, while the reversed expanding motion is repulsive, exponentially unstable, and unobservable.
{\bf Figure 10} illustrates cross sections of an attractor-repellor pair for the heat
conducting oscillator of Section VII with maximum temperature gradient $\epsilon = 0.50$ .
Neither multifractal object, the attractor nor its mirror-image repellor, with $\{ \ +p
\longleftrightarrow -p \ \}$ exhibits topological symmetry.  Of the 183,264 attractor points
shown in the figure 117,364 correspond to positive Lyapunov exponents and 65,900 to negative.
It would be worthwhile to carry out detailed comparisons of the Lyapunov exponents both forward
and backward in time for such problems.

Thw preponderance of positive local exponents, $\lambda_1(t)$ , is typical of time-reversible
chaotic systems either at, or away from, equilibrium.  In the nonequilibrium case the smooth Gibbsian
phase-space probability density is replaced by the ``strange'' attractive flow, signalling
the loss of phase volume to dissipation.  There is a topological symmetry breaking with the
two phase-space structures, attractor and repellor, both obeying identical equations of
motion, while one has measure zero and the other has {\it all} the measure, unity.

\noindent

\begin{figure}
\includegraphics[width=4.5in,angle=+90.]{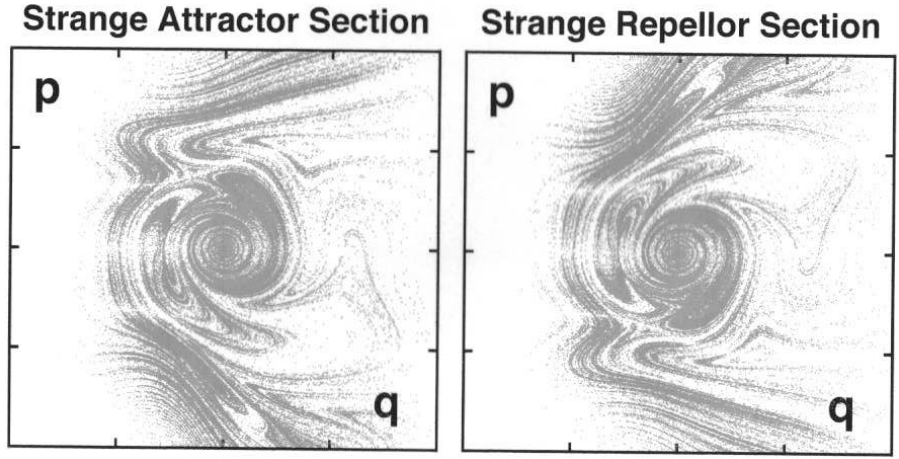}
\caption{
The nonequilibrium harmonic oscillator, with $T(q) = 1 + 0.50\tanh(q)$ , shows no Lyapunov
symmetry whatever in its $(qp0)$ cross section. Just as at equilibrium the dynamics can only
depend on the past --- there is no symmetry on changing the sign of $p$ .  At the same time
the dynamics must obey the Second Law of Thermodynamics --- the strange attractor section
( at the left ) corresponds to a mean current $\langle \ (p^3/2) \ \rangle < 0$ .  The strange
repellor section ( at the right ) is unobservable, with a positive current and a positive 
time-averaged Lyapunov-exponent sum.  The local attractor and repellor Lyapunov exponents are
not at all simply related.  The overall flow in phase space is ``from'' the fractal repellor
``to'' the attractor. The mirror-image sections shown here correspond to 183,264 penetrations
of the $\zeta = 0$ plane.  The two sections {\it are} mirror images.
}
\end{figure}

The Galton Board problem ( a particle falling at constant kinetic energy through a periodic
array of scatterers ) and the conducting oscillator problem ( an oscillator conducting heat in
the presence of a temperature gradient ) both furnish three-dimensional models of nonequilibrium
steady states. Both provide ergodic fractal geometry, Lyapunov instability, and the symmetry
breaking associated with time reversal and with the Second Law of Thermodynamics.  This explanation
of the Second Law is identical for manybody problems too.  My interest in few-body systems was
inherited from my earliest work with Berni, dating back to our exploration of one-body cell models
for melting as well as our subsequent few-body thermodynamic studies of disks and spheres.

The Lyapunov {\it spectrum} ( with all $n$ exponents describing the comoving expansion and contraction
in an $n$-dimensional space ) characterizes the spatial dependence of instabilities in {\it all}
$n$ of the phase-space directions.  It shows that the predominantly negative Lyapunov spectrum
describing the condensation of attractive sets onto fractal objects mirrors the predominantly positive
spectrum describing the exponentially fast departure of trajectories in the vicinity of the
repellor.  It is interesting and significant that despite the time-reversibility of the equations of
motion the stability of the motions forward and backward is qualitatively different.  Just as on a
roller coaster or a curvy road the passengers' motions are sensitive to both the direction of travel
{\it and} the strange-attractor direction imposed by the Second Law.

Caricatures of these problems, carried out here at Livermore, and at Wien, led to this new
understanding of the Second Law of Thermodynamics, first described in a 1987 paper ( submitted twice,
with two different titles ! ) with Brad Holian and Harald Posch\cite{b21}.
In nonequilibrium steady states the motion forward in time has an attractive fractal distribution
in phase space with a phase volume of zero.  The vanishing volume corresponds to the extreme rarity
of the states participating in a steady flow.  The time-reversed states, equally rare, make up an
inaccessible repellor with a summed-up Lyapunov spectrum which is positive rather than negative. In
this way we found that a {\it time-reversible} mechanics, based on Nos\'e's ideas, provides a clear
foundation for {\it irreversible} thermodynamics.

\section{Conclusion}
Berni's research style, beginning with a simple model confirming a ``horseback guess'', followed by
painstaking analyses leading to a clear intuitive description of the work and its significance,
has led to rapid progress in understanding phase transformations, nonlinear transport, and aspects
of dynamical-systems theory.  The 2013 Nobel Prizes in Chemistry rewarded Martin Karplus, Michael
Levitt, and Arieh Warshel for applications of {\it classical} and {\it thermostated} molecular dynamics
to biomedical problems.  Berni has always emphasized the need for simplicity and clarity in his 
work, with an
emphasis on words rather than equations and intuitive arguments rather than formal proofs.  After
more than a half century of research his way of working seems natural to me and I recognize Berni
as a good part of its source. I am looking forward to seeing more of his inspirational work in the
years ahead.

\section {Acknowledgments}

My Wife and colleague Carol has helped me throughout.  She, Berni, and hundreds of collaborators
and colleagues have my Love and Gratitude.  I particularly appreciated Brenda Rubenstein's
tireless organizational skills and the staff at the Lawrence Livermore National Laboratory for
helping Carol and me to attend and enjoy Berni's Birthday Symposium.

\pagebreak

\end{document}